\documentstyle[aas2pp4,tighten]{article}  

\def\nice#1::::{#1}    \def\subm#1::::{}   

\subm \documentstyle[12pt,aasms4]{article} ::::

\nice \def\apjsmall{\small} ::::
\subm \def\apjsmall{} ::::

\nice \textheight 260mm \headheight 0pt \topmargin -2.1cm ::::

\def\k{\,km\,s$^{-1}$}
\def\centreline{\centerline}
\def\.{{\cdot}} 
\def\gtapprox{\,\lower.6ex\hbox{$\buildrel >\over \sim$} \, }
\def\ltapprox{\,\lower.6ex\hbox{$\buildrel <\over \sim$} \, }
\def\propapprox{\,\lower.6ex\hbox{$\buildrel \propto\over \sim$} \, }

\def\e{ {\scriptstyle \times} 10^}

\def\arcs{\ifmmode {'' }\else $'' $\fi}     
\def\arcm{\ifmmode {' }\else $' $\fi}       
\def\deg{\ifmmode^\circ\else$^\circ$\fi}    

\def\fr7{7$ \hskip -0.9ex \vrule height0.8ex width0.8ex depth-0.73ex
                                                                \hskip0.1ex$}
\def\frtoday{Le\space\number\day\space\ifcase\month\or
  janvier\or f\'evrier\or mars\or avril\or mai\or juin\or
  juillet\or ao\^ut\or septembre\or octobre\or novembre\or d\'ecembre\fi\space \number\year}

\newcommand\joref[5]{#1, #5, {#2, }{#3, } #4}  
\newcommand\confref[5]{#1, #5, {#2, }{#3, } #4.} 
\newcommand\inpress[5]{#1, #5, #2, in press}  
 
\newcommand\epref[3]{#1, #3, #2}

\def\reff{$r_{\mbox{\rm \small eff}}^p$}
\def\reffninety{$r_{\mbox{\rm \small eff}}^{90\%}$}
\def\refffifty{$r_{\mbox{\rm \small eff}}^{50\%}$}
\def\zmed{$z_{\mbox{\rm \small med}}$}
\def\dprop{d_{\mbox{\rm \small prop}}}

\slugcomment{Accepted for publication in The Astrophysical Journal}

\lefthead{Roukema \& Valls-Gabaud}
\righthead{Spatial separations of galaxies in deep surveys}

\begin{document}

\title{On the Effective Spatial Separations\\ 
in the Clustering of Faint Galaxies}
\author{Boudewijn F. Roukema}
\affil{Division of Theoretical Astrophysics,
      National Astronomical Observatory, Mitaka, Tokyo 181, Japan\\
	\nice {\rm Email: } {\tt roukema@iap.fr} ::::
	\subm Email: roukema@iap.fr :::: }
\and 
\author{David Valls-Gabaud}
\affil{URA CNRS 1280, Observatoire de Strasbourg, 
                 11 rue de l'Universit\'e, 67000 Strasbourg, France\\
       Royal Greenwich Observatory, Madingley Road,
                 Cambridge CB3 0EZ, UK\\
\nice       {\rm Email: } {\tt dvg@astro.u-strasbg.fr  dvg@ast.cam.ac.uk} ::::
\subm      Email:  dvg@astro.u-strasbg.fr  dvg@ast.cam.ac.uk :::: }

\author{{\small Accepted for publication in The Astrophysical Journal} }

\begin{abstract}
Several recent measurements have been made of the angular correlation function
$w(\theta,m)$
of faint galaxies in deep surveys (e.g., in the Hubble Deep Field, HDF).
Are the measured correlations indicative of gravitational growth of
primordial perturbations or of the relationship between galaxies and
(dark matter dominated) galaxy haloes? A first step in answering this 
question is to determine 
the typical spatial separations of galaxies whose {\em spatial} correlations,
$\xi(r,z),$ 
contribute most of the angular correlation.

The median spatial separation of galaxy pairs contributing to 
a fraction $p$ of the angular correlation signal in a galaxy survey 
is denoted by {\reff} (\S\protect\ref{s-reff}) and compared with
the perpendicular distance, 
$r_\perp,$ at
the median redshift, {\zmed}, of the galaxies. 
Over a wide range in spatial correlation growth rates $\epsilon$ and
median redshifts, 
{\refffifty} is no more than
about twice the value of $r_\perp,$ while 
{\reffninety} is typically four times $r_\perp.$ 

Values of {\reff} for redshift distributions representative 
of recent surveys indicate
that many angular correlation measurements correspond to spatial 
correlations at comoving length scales well below 1$h^{-1}$~Mpc. 
For $\Omega_0=1$ and $\lambda_0=0$, the correlation signal at {4\arcsec}
predominant in the \cite{Vill96}~(1996) 
estimates of $w(\theta,m)$ for faint HDF galaxies
corresponds to {\refffifty}~(HDF)~$\approx 40h^{-1}$~kpc; other cosmologies and 
angles up to {10\arcsec} can increase this to 
{\refffifty}~(HDF)~$\ltapprox 200h^{-1}$~kpc. The proper separations are
$(1+z)$ times smaller, where $1\ltapprox z \ltapprox 2.$

These scales are small: the 
faint galaxy angular correlation measurements are at scales where
halo and/or galaxy existence, let alone interactions, may 
modify the spatial correlation function. These measurements could be
used to probe the radial extent of haloes at high redshift.

\end{abstract}

\keywords{cosmology: theory---galaxies: formation---galaxies: 
clusters: general---galaxies: distribution---cosmology: observations}

\def\fperc{ 
\begin{figure}
\centering 
\nice \centreline{\epsfxsize=8cm
 \epsfbox[60 38 493 739]{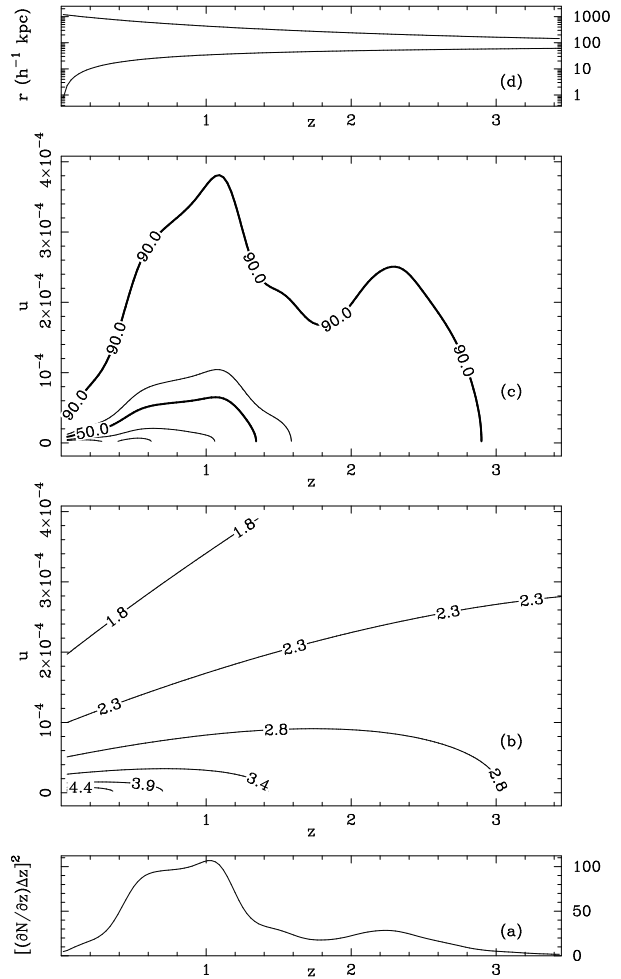}} ::::
\caption[Percentage of $w_0$ contours.]{ \apjsmall
\label{f-perc} Analysis of how the spatial correlations $\xi$ combine
with the number distribution $\partial N/\partial z$ to form the
integral $w(\theta,m).$ Each of the panels is plotted against the same
redshift range horizontally. From bottom to top, the four panels show:
(a) $[N_z(z,m) \Delta z]^2$ in units of (squared) observed numbers of galaxies
in bins of $0\.2$ $z$ units;
(b) contours of 
$\log_{10}\{\xi[r(z,u),z]\}$ in the $z-u$ plane, showing the highest values
at low $z$ and low $u$;
(c) contours of constant $(\partial N/\partial z)^2 \xi[r(z,u),z]$ 
in the $z-u$ plane within which a fraction $p$ 
of the numerator of Eq.~\protect\ref{e-limber} is obtained, for 
$p=10\%, 30\%, 70\%$ (thin curves) and $p=50\%, 90\%$ (thick curves);
(d) comoving separations (in $h^{-1}$~kpc) 
corresponding to the $u$ range plotted 
in (b),(c)---$\log_{10}r(z,u=0)$ and $\log_{10}r(z,u=4\.1\e{-4})$ are the
lower and upper curves respectively. The redshift distribution is 
the HDF photometric $z$ distribution of \protect\cite{Sawi97}~(1997).
(The published histogram is smoothed by a gaussian of width 
$\sigma_z=0\.1.$)
Parameter values are $\Omega_0=1,$ $\lambda_0=0,$ $\theta=4\arcs,$ 
$r_0=5\.5 h^{-1}$~Mpc, $\gamma=1\.8,$ $\epsilon=0$.
}
\end{figure} } 

\def\fepsz0{ \begin{figure}
\nice 
\epsfxsize=8cm
 \epsfbox[52 39 501 601]{"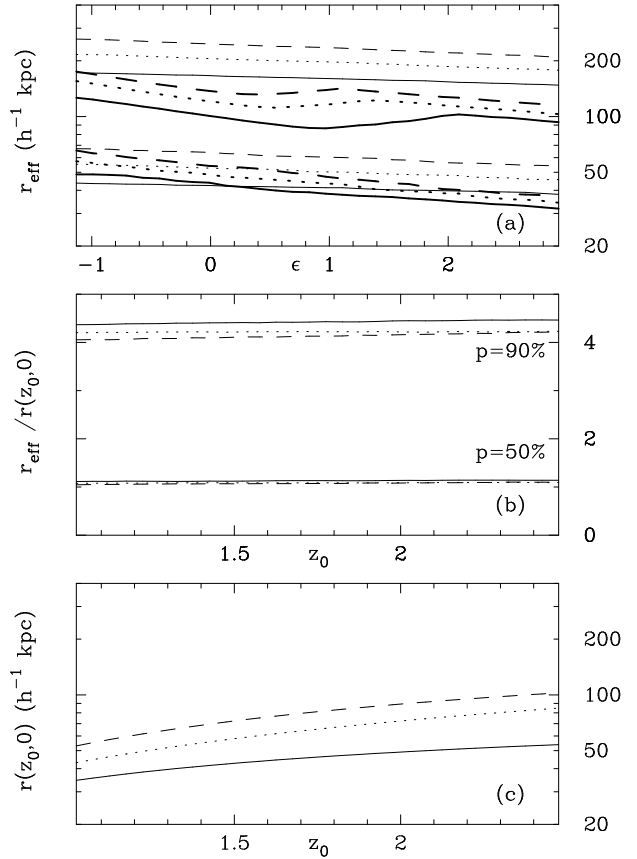"} 
 ::::
\caption[Dependence on $\epsilon$ and $z_0$]{ \apjsmall
\label{f-epsz0} (a) (top panel) Dependence of \protect{\reff} [the 
median (comoving) radius 
which contributes a fraction 
$p$ of the double integral determining $w(\theta,m)$] 
on the growth rate $\epsilon$
of the spatial correlation function.
The upper (lower) six curves are for $p=90\%;$ ($p=50\%$).
Curves are 
thick for \protect\cite{Sawi97}'s (1997) HDF $I_{AB,8140}<27$ photometric
redshift distribution and 
thin for the analytical distribution 
of Eq.~\protect\ref{e-smoodndz} with $z_0=1\.20$ (having the same 
\protect{\zmed} as \protect\cite{Sawi97}'s redshift distribution).
Different cosmological models 
are shown [in (a)-(c)] by solid ($\Omega_0=1, \lambda_0=0$), 
dotted ($\Omega_0=0\.1, \lambda_0=0$) and 
dashed ($\Omega_0=0\.1, \lambda_0=0\.9$) curves. 
(b) (middle panel) Dependence of \protect{\reff} 
on $z_0,$ adopting 
Eq.~\protect\ref{e-smoodndz} 
(for which $z_0\approx 0\.95$~\protect{\zmed}), 
shown divided by minimum perpendicular distance at a given
redshift $r_\perp\equiv r(z_0,0).$ 
Upper (lower) three curves are for $p=90\%;$ ($p=50\%$).
(c) (bottom panel) 
Dependence of $r_\perp\equiv r(z_0,0)$ on $z_0$ (equivalent to the
angular diameter-redshift relation).
Parameter values held constant in all curves are 
$\theta=4$\arcs, $r_0=5\.5 h^{-1}$~Mpc and $\gamma=1\.8.$ 
}
\end{figure} 
} 

\def\fshape{ 
\begin{figure}
\nice 
\epsfxsize=8cm
 \epsfbox[52 39 501 631]{"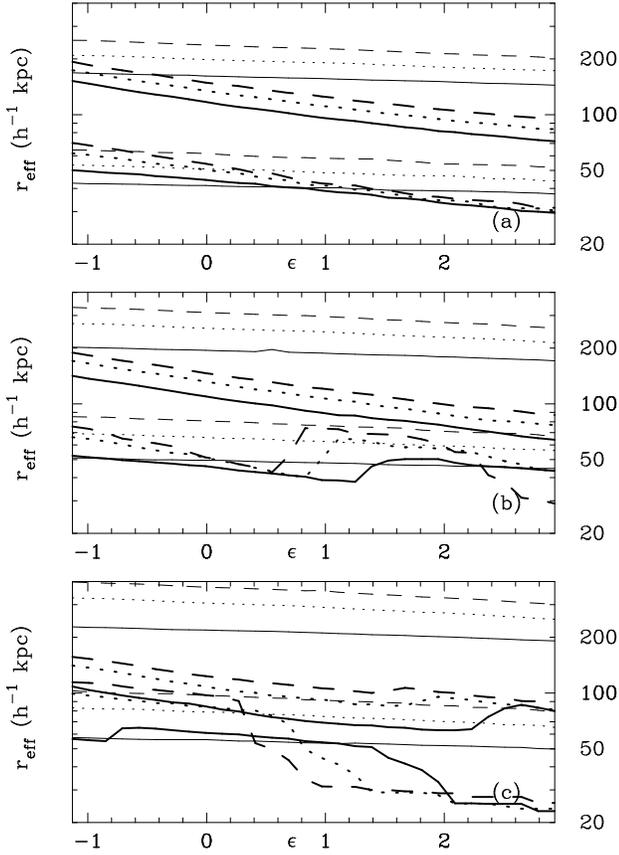"} 
 ::::
\caption[Dependence on Shape of dN/dz]{ \apjsmall
\label{f-shape} 
The dependence of \protect{\reff} on the shape of $N_z$ is 
shown by reproducing Fig.~\protect\ref{f-epsz0}(a), but 
replacing the $N_z$ estimate of 
\protect\cite{Sawi97}~(1997) by those of  
(a) \protect\cite{Lanz96}~(1996), 
(b) \protect\cite{Mob96}~(1996, Fig.2) and 
(c) \protect\cite{GwHart96}~(1996), 
and using analytical distributions with appropriate $z_0$ 
values (a) $z_0=1\.15$, (b) $z_0=1\.6$ and (c) $z_0=2\.0.$
Axes and line styles are as for Fig.~\protect\ref{f-epsz0}(a).
}
\end{figure} 
} 

\def\fdndz{ 
\begin{figure}
\nice 
\epsfxsize=9cm
 \epsfbox[77 39 526 321]{"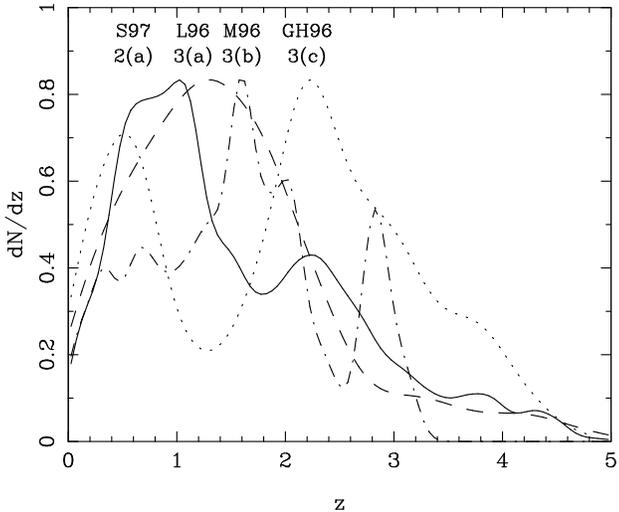"} 
 ::::
\caption[HDF Redshift Distributions]{ \apjsmall
\label{f-dNdzHDF} 
The shapes of the photometric redshift distributions $N_z$ estimated by 
\protect\cite{Sawi97}~(1997), \protect\cite{Lanz96}~(1996), 
\protect\cite{Mob96}~(1996) and \protect\cite{GwHart96}~(1996)
for the HDF are shown here against redshift, normalised for 
purposes of comparison. The dependence of \protect{\reff} on the 
different shapes is shown in Figs 2(a), 3(a), 3(b), 3(c) as 
indicated.
}
\end{figure} 
} 

\def\tabColl{ 
\begin{table}
\caption{\label{t-Colley}
Values of {\protect\reff} for $1\arcsec$ for redshifts 
$z=2\.5$ and $z=5$ spanning redshifts likely
for the HDF objects colour-selected to high redshift by
\protect\cite{Colley96}~(1996; 1997), assuming that 
Eqs~\protect\ref{e-reff}-\protect\ref{e-reffdel} can be validly extrapolated
to these objects. These effective separations are in units of 
$h^{-1}$~kpc, comoving or proper as indicated.}
\begin{tabular}{c cc cccc}
\hline \multicolumn{1}{c}{$p$} &
\multicolumn{2}{c}{cosm.} &
\multicolumn{2}{c}{comoving}  & \multicolumn{2}{c}{proper}\\ 
\hline
 & $\Omega_0$ & $\lambda_0$ &  $z=2\.5$ & $z=5$ & $z=2\.5$ & $z=5$  \\
\hline
50\%  &    $1\.0$ &     $0\.0$ &    14 &    17 &     4 &     3 \\ 
50\%  &    $0\.1$ &     $0\.0$ &    21 &    35 &     6 &     6 \\ 
50\%  &    $0\.1$ &     $0\.9$ &    26 &    37 &     7 &     6 \\ 
90\%  &    $1\.0$ &     $0\.0$ &    54 &    69 &    16 &    12 \\ 
90\%  &    $0\.1$ &     $0\.0$ &    86 &   141 &    24 &    24 \\ 
90\%  &    $0\.1$ &     $0\.9$ &   103 &   147 &    30 &    25 \\
\hline
\end{tabular}
\end{table}
   } 


\section{INTRODUCTION}
The growth of primordial density fluctuations 
into galaxies and clusters of 
galaxies is an essential element of understanding how structure forms in 
the Universe. A common way to statistically represent this structure 
from observed galaxies is to calculate the two-point auto-correlation 
function, $\xi(r,z),$ which can be approximately parametrised 
as a power law (in 
spatial separation of galaxy pairs) which increases in amplitude 
as a function of time,  
according to a single parameter $\epsilon,$ 
(e.g., \cite{GroP77}~1977):
\begin{equation}
\xi(r,z) = (r_0/r)^{\gamma} (1+z)^{-(3+\epsilon-\gamma)}
\label{e-defneps}
\end{equation}
where $r$ and $r_0$ are expressed in comoving coordinates and 
$\gamma$ represents the approach to homogeneity at larger length scales.

Observed values are typically $r_0\approx5h^{-1}$~Mpc 
and $\gamma\approx 1\.7-1\.8$ (e.g., \cite{DavP83}~1983; 
\cite{Love92}~1992) for the low redshift general galaxy population. 
\cite{DavP83}' (1983) analysis is consistent with this power law on scales 
$10h^{-1}$~kpc$ \ltapprox r \ltapprox 10h^{-1}$~Mpc, though the correlations
appear somewhat stronger at the small scale end (but noisy). 
\footnote{The more recent estimate of \cite{Tuck97}~(1997) finds similar behaviour
on scales down to about $20h^{-1}$~kpc, but for the redshift-space
correlation function rather than the ``real'' space correlation function,
which makes this difficult to interpret.}

\cite{Love92} (1992) find their power law fit to $\xi$ of the Stromlo-APM redshift 
survey to be valid over $200h^{-1}$~kpc$ \ltapprox r \ltapprox 20h^{-1}$~Mpc.

Direct estimates of the evolution in this power law at low {\zmed} 
(median redshifts) include 
$\epsilon=1\.6\pm0\.5$ (\cite{WIHS93}~1993, 
{\zmed}$=0\.4$), 
$\epsilon=-2\.0\pm2\.7$ 
(\cite{CEBC}~1994, {\zmed}$=0\.16$)
\footnote{\protect\cite{CEBC}'s (1994) $\varepsilon$ relates to
$\epsilon$ as $\epsilon=\gamma( 1- \varepsilon ) -3.$}
and $\epsilon=0\.8^{+1.0}_{-1.3}$ 
(\cite{Shep96}~1996, {\zmed}$=0\.36$). However, 
consideration of galaxies as
a single population, i.e., adoption of 
$r_0=5\.0 h^{-1}$~Mpc and the 
median redshift ($c${\zmed}$=15,200$\k) 
 of the Stromlo-APM survey (\cite{Love92}~1992, 1995) 
for comparison with the higher {\zmed}$=0\.56$ CFRS estimate of $\xi$ 
(Le~F\`evre et al. 1996, \cite{CFRS-VIII})
would imply that $\epsilon=2\.8$ (cf. \S 4.1(3), \cite{CFRS-VIII}). 

This latter value of $\epsilon$ is considerably higher than expected 
either for clustering fixed in comoving coordinates 
($\epsilon=\gamma-3 \approx -1\.2$, 
so that the index in Eq.~\ref{e-defneps} is zero); 
clustering fixed in proper coordinates on small
scales [``stable clustering'', $\epsilon=0$, since the numbers of 
``clusters'' changes by $(1+z)^3$ and the number of galaxy pairs 
by $(1+z)^6$, the factor in Eq.~\ref{e-defneps} for $r$ in proper 
coordinates is $(1+z)^{-3}$]; or linear growth of density perturbations
in an Einstein-de Sitter universe ($\epsilon=\gamma-1 \approx 0\.8$).  
[Note also that the high $z$ behaviour of $\xi$ may be different (e.g., 
\cite{ORY97}~1997).]

A lower value of $\epsilon$ could be justified by hypothesising major
changes in visible galaxy populations from low to high redshifts, 
as suggested by indications of colour dependence in the correlation 
function (e.g., \cite{IP95}~1995).

The validity or otherwise of the high observational 
estimates of $\epsilon$ or of major changes in the general galaxy 
population is not
the subject of this article. In fact, as will be defined and 
seen below, the parameter of interest {\reff} is only weakly sensitive
to $\epsilon.$ 

The motivation for this article stems from the attempts to indirectly
measure the dependence of $\xi$ on $z.$
Since estimation of spectroscopic redshifts requires many more photons 
than the photometric detection of a galaxy, 
the projection of $\xi$ onto 
the celestial sphere, i.e., the angular correlation function, $w(\theta,m),$ 
for angle $\theta$ and survey ``limiting apparent magnitude'' $m,$ 
can be more easily measured than $\xi.$ This means that $\xi$ can be
effectively measured at larger redshifts than is obtainable in redshift
surveys---but at the cost of having to deduce $\xi$ from its sky 
projection.

Hence, investigation of the $z$ dependence of $\xi$ can be carried out
to larger redshifts by measuring $w$ rather than $\xi.$
Many estimates of the amplitude of $w(\theta,m)$
of faint galaxies in faint magnitude (``deep'') small angle 
(a few square arcminutes, ``pencil beam'', or up to a few square degrees)
surveys have been carried out recently
(e.g., \cite{Ef91}~1991; 
\cite{Neu}~1991; \cite{PI92}~1992; \cite{Couch93}~1993;
\cite{RP94}~1994; \cite{IP95}~1995; \cite{BSM95}~1995). 
These angular correlations are usually interpreted by integrating
$\xi(r,z)$ (Eq.~\ref{e-defneps}) 
over an appropriate domain in $z-u$ space (see Eq.~\ref{e-limber}). 

The resulting values of $\epsilon,$ or hypotheses about how population
transitions might justify change in the value of $r_0$ as a function
of $z,$ are generally discussed, and an indication of the scale 
of galaxy pair {\em separations} is sometimes indicated by 
 $r_\perp\equiv \dprop \theta \equiv r(z,u=0)$ 
(for $r$ in comoving coordinates; $u$ is redshift separation of 
a galaxy pair at mean redshift $z$, defined 
below for Eq.~\ref{e-limber}) at
the median redshift, {\zmed}, of the redshift distribution of the
galaxy sample. 

However, the integral also includes galaxy pairs at
unequal redshifts ($u\neq 0$) and galaxy pairs at mean redshifts lower and
higher than {\zmed}, so rather than 
the usual assumption that $r_\perp$ is representative of
typical scales contributing to $w,$ it is preferable
to analyse the integral more closely. 

This is the goal of this article: to determine what pair separations
contribute to $w(\theta,m)$ over a region of parameter space judged
likely for observational values of $w(\theta,m).$
This is done by (a) separating out the different contributions in 
the double integral relating $\xi$ to $w,$ and by (b) defining an 
effective separation {\reff} to be the median separation of galaxy 
pairs which contribute to a fraction $p$ of the numerator of this integral. 
The effective separation depends, in principle, on
both the redshift distribution and the evolution of $\xi,$ 
so {\reff} is evaluated for likely ranges of relevant parameters.

Indeed, the assumption that $r_\perp$ is a typical scale of galaxy pairs
contributing to $w$ turns out to be a good intuition, to better than
an order of magnitude. In this paper we present quantitative 
justification for this intuition.

In \S\ref{s-lim}, the double integration of $\xi(r,z)$ is presented, 
using \cite{Sawi97}'s~(1997) photometric redshift distribution for 
the Hubble Deep Field (HDF; \cite{Will96}~1996) as an illustration.
The effective separation, {\reff} 
is defined and evaluated in \S\ref{s-reff}.
In \S\ref{s-discuss}, 
implications of the resulting {\reff} values are discussed 
and \S\ref{s-conclu} presents the conclusions. All discussion
is in comoving ($t=t_0$) units unless otherwise noted and the Hubble 
constant is $100h$~\k~Mpc$^{-1}$.

\section{LIMBER'S EQUATION}\label{s-lim}
The angular correlation function (small angle approximation) 
is given by the the double integration of $\xi(r,z),$ 
\begin{equation} w(\theta,m) = 
{ \int dz\, N_z(z,m)^2 \int du\, \xi(r,z)
	\over
 \left[\, \int dz\, N_z(z,m) \,\right]^2}
\label{e-limber}
\end{equation}
where $\theta$ is the angle on the sky, $m$ is the apparent magnitude, 
$z=(z_1+z_2)/2$ and $u=z_1-z_2$ parametrise the redshifts of two galaxies
at redshifts $z_1$ and $z_2,$ $r(z,u)$ is the spatial separation between
the two galaxies, and 
$N_{z}(z,m)=\partial^2 N/\partial m \partial z $ is the redshift 
distribution at $m$ (\cite{Lim53}~1953; \cite{Ph78}~1978; \cite{Bible}~1980;
\cite{Ef91}~1991). 

\nice\fperc::::

One can think of $\xi(r,z)$ as the excess probability that two randomly
chosen galaxies lie at a (comoving) separation $r.$ By symmetry, 
all such galaxies
separated in projection by $\theta$ can be thought to form 
 the surface of a cone of opening angle $\theta/2.$ Again by symmetry,
the dependence on the second angle $\phi$ can be dropped, 
so that integration is only needed over the
double range of redshifts of pairs of galaxies, weighted by the 
numbers of galaxies at each of the two redshifts. 
Note that the angular correlation function $w$ is independent of 
the normalisation of $\partial^2 N/ (\partial z\partial m),$ so only
its shape is relevant to the integral (e.g. \cite{Yosh93} 1993).

In order to understand the relative importance of the different  
factors in the integral, 
we separate them out in 
Fig.~\ref{f-perc}. The redshift distribution chosen for this illustration
is \cite{Sawi97}'s~(1997) photometric redshift distribution for 
the HDF galaxies (see also \cite{Lanz96}~1996) in the three Wide Field Camera images of the 
Version 2 (1996 February 29) drizzled version 
of the data to 
$I_{AB,8140}<27$. 
The square of this distribution is
shown in Fig.~\ref{f-perc}(a).

The redshift and distance dependence of $\xi[r(z,u),z]$ can be seen in 
Fig.~\ref{f-perc}(b). 
At any given $z,$ $\xi$ 
decreases as $u$ increases, since the spatial correlation decreases
for increasing $r$, quickly approaching zero. Both the use of a fixed
angle $\theta$ (chosen here as $\theta=4\arcsec$, 
since the range of significant signal
in \cite{Vill96}'s $w_0$ HDF estimate is roughly 
$2\.5\arcsec$ to $10\arcsec$) 
and the choice of 
``stable clustering in proper coordinates'' (i.e., $\epsilon=0$) 
contribute to the decrease in $\xi$ along the path for which $u\equiv 0.$
For large (fixed) $u,$ 
$r(z,u)$ becomes dominated by a nearly radial distance
separation. Since $d^2 \dprop /dz^2 < 0,$ these separations
become smaller for increasing $z,$ which would imply higher $\xi$ for
clustering fixed in comoving coordinates ($\epsilon=\gamma-3$) as 
$z$ increases. Indeed, Fig.~\ref{f-perc}(b) shows that for ``large'' constant 
values of $u,$ the increase in $\xi$ as $z$ increases is strong enough
to overcome the weakening of the amplitude of $\xi$ for higher $z.$

Fig.~\ref{f-perc}(c) shows that neither one nor the other of these limits
in $u$ is sufficient to cover the domain in $z-u$ space which is relevant
for the total value of the double integral. The maximum values of $u$
necessary to cover 90\% of the numerator 
in Eq.~\ref{e-limber} are less than about $0\.05\%$ 
of a redshift ``unit'', but this is still large enough that the dominance
of radial separation over perpendicular separation causes $\xi$ to 
increase as $z$ increases at a fixed $u,$ at least for $\theta=4\arcsec.$

The smallness of the value of $u$ required to account for even 50\% of
the numerator of Eq.~\ref{e-limber} 
is the key element required to interpret observational
estimates of $w(\theta,m).$
 In Fig.~\ref{f-perc}(d), the $u$ range covered in the
$z-u$ plane plots is shown in units of (comoving) $h^{-1}$~kpc. 
Nearly all of the 90\% contour is for galaxies separated by less than 
 $1 h^{-1}$~Mpc; much of the integral comes from separations less
than  $100 h^{-1}$~kpc. This is the scale on which galaxy interactions
and the relationship between galaxy haloes dominated by dark matter 
and the visible galaxies containing stars are likely to be complicated.

Other points to be noted in Fig.~\ref{f-perc}(c) are:

(1) \cite{Sawi97}'s (1997) redshift distribution includes many galaxies
in the lowest redshift bins. Whether this is due to a genuinely steep
faint end of the galaxy luminosity function or to the use of photometric
rather than spectroscopic redshifts, it is not of cosmological interest
to include spatial correlations of local dwarf galaxies 
within a few hundred
kiloparsecs of the observer in estimates of $w(${\zmed}$)\gtapprox 1.$
It is therefore of importance to observe that 
the percentile contours concentrate towards lower and lower $z$ for
higher values of the integrand (lower percentiles). The combination
of rapidly decreasing separations, increasing values of $\xi(r,z)$
and slowness of $N_z(z,m)$ to decrease imply that a significant fraction
of the signal in $w(\theta,m)$ could be due to very local galaxies.

For calculations presented below,  
a (conservative) lower limit is therefore set 
such that redshifts for which the perpendicular 
separation $r(z,0)$ is less than $10h^{-1}$~kpc
 are excluded from the domain of integration.

(2) \cite{Sawi97}'s (1997) $N_z$ peak at $z\approx 2\.3$ contributes significantly
between the $70\%$ and $90\%$ contours, i.e., roughly $10\%$ of the 
numerator in Eq.~\ref{e-limber} is due to this peak. Given the 
uncertainties 
in \cite{Vill96}'s (1996) estimate of the HDF angular correlation
function 
(essentially Poisson due to the small numbers of objects),
this is not likely to be important for interpretation of the HDF data.
However, for the brighter and more precise 
angular correlation function estimates (e.g., \cite{BSM95}~1995)
this could be more important---analytical single-peaked redshift distributions
may not be precise enough.

\nice\fepsz0::::

\section{THE EFFECTIVE SEPARATION \protect{\reff}}
\label{s-reff}

Of course,
the values of the separation scales which contribute most of the value
of $w(\theta,m)$ discussed above are for a particular choice of
the parameters chosen in 
Eq.~\ref{e-defneps} (or on the validity of this equation at such
scales), the redshift distribution, $\theta$ and the cosmological model.
In order to discuss dependence of the separation scales on
Eq.~\ref{e-defneps},  
on the {\zmed} and the shape of the redshift distribution, 
on $\theta$
 and on cosmological parameters, it is useful
to define the {\em effective separation,\/} {\reff}, to be the median value of 
$r(z,u)$ over the domain in $z-u$ space inside a contour of 
constant $N_z^2(z,m) \xi[r(z,u),z]$ 
which contributes a fraction $p$ of the value of $w(\theta,m)$ 
(where the denominator in Eq.~\ref{e-limber} is held fixed).

\subsection{Dependence on the Spatial Correlation Function} \label{s-xidep}

Lowering the value of 
$r_0$ (which is considered by, e.g., \cite{Bernst94}~1994; 
\cite{BSM95}~1995, to denote a possible transition in galaxy ``populations'')
factors out of Eq.~\ref{e-limber} as a constant; it reduces the value of
$w$ but does not affect the question of which separations are relevant to
the integral. Reducing the value of $\gamma$ (for fixed $\epsilon-\gamma$ in
order not to affect the redshift evolution of $\xi$) would increase the
values of {\reff} slightly. 
In estimates of $w$ for faint galaxies, it is usually 
assumed that $\gamma - 1=0\.8$ in order to correct for 
the integral constraint, so this is the value of interest here
(e.g., \cite{HudLil}~1996; though note \cite{Camp95}~1995; \cite{IP95}~1995).

Of more interest is the value of $\epsilon,$ which as
mentioned above has either theoretically or observationally motivated 
values lying in the range from $-1\.2$ to nearly $3.$

Fig.~\ref{f-epsz0}(a) 
shows that in spite of the wide range in values of $\epsilon,$ 
the dependence of {\reff} on $\epsilon$ is weak. 
Even for a cosmological constant 
dominated flat cosmological model and $\epsilon=3,$
{\reff} for \cite{Sawi97}'s (1997) redshift distribution is no more than
a factor of two lower than that for the analytical distribution.

Although weak, there is an average trend to
lower values for higher $\epsilon.$ Higher $\epsilon$ means a 
more rapid decrease in spatial correlation amplitude with $z,$ 
so that the correlations at lower $z$ contribute relatively more to 
the integral. The lower the value of $z,$
the lower the (comoving) 
perpendicular separations $r(z,0),$ for any cosmological model.
Hence, the slight decrease in {\reff}. 

In other words, {\em independently of the growth rate of the spatial 
correlation function}, angular correlations in deep surveys of faint 
galaxies only probe length scales characteristic of galactic haloes.

The stronger decrease in {\reff} for the observational (photometric)
redshift distribution than for the analytical distribution of the 
same {\zmed} can be attributed to its complex shape. Once $\epsilon$
has increased enough that the $z\approx 2\.2$ peak no longer contributes
much, the effective {\zmed} of the observational distribution will be
lower than that of the analytical distribution; thus, the lower {\reff}
values. 

The trend to lower {\reff} is not monotonic for the observational 
redshift distribution. This is attributable to an increase in the
relative importance of $u$ separations to perpendicular separations
when $N_z^2(z,m)(1+z)^{-(3+\epsilon-\gamma)}$ becomes flatter.
A ``complex'' enough redshift distribution could possibly increase
the importance of this reversed dependence of {\reff} on $\epsilon,$ 
as indeed is seen below.

\subsection{Dependence on the Redshift Distribution} \label{s-dndzdep}

\nice\fshape::::
\nice\fdndz::::

The effective separation {\reff} depends on a redshift distribution 
primarily via 
the typical redshifts in the sample, or more quantitatively, 
on the median redshift, {\zmed}, of $N_z.$ This dependence is analysed here 
by use of the simple analytical distribution 
\begin{equation}
z^2 \exp[-(z/z_0)^\beta], 
\label{e-smoodndz}
\end{equation} 
where $\beta=2\.5$ (as in \cite{Ef91}~1991; \cite{Vill96}~1996). 
For this distribution, {\zmed} dependence translates directly
into dependence on $z_0,$ since
{\zmed}~$\approx 0.95 z_0$. 

A very useful property of this dependence is shown in 
Fig.~\ref{f-epsz0}(b): nearly all of this $z_0$ dependence is in the
(equivalent of the) angular diameter-redshift relation, shown as the
dependence of the perpendicular separation $r_\perp(z)\equiv r(z,u=0)$ on $z$
in Fig.~\ref{f-epsz0}(c).

Both {\refffifty} and {\reffninety} are nearly constant over a wide
range in $z_0.$ Moreover, {\refffifty} is only about 
$5\%-15\%$ greater than $r_\perp(z_0),$ i.e., about $10\%-20\%$ 
greater than $r_\perp(${\zmed}$).$ Relative to the precision of present
cosmological measurements, it can be said that $r_\perp(${\zmed}$)$ 
is a very good estimator for {\refffifty} for such analytical
distributions. In addition, since {\refffifty} is 
insensitive to $\epsilon$ and since the 
$\epsilon$ dependence of both the observational 
(photometric) and analytical $N_z$ curves are similar 
(Fig.~\ref{f-epsz0}), 
the quality of $r_\perp(${\zmed}$)$ as an estimator for {\refffifty}
is little dependent on the value of $\epsilon$ and the precise shape 
of $N_z,$ 
 avoiding the need for explicit integration of 
Limber's Equation (Eq.~\ref{e-limber}).

The $90\%$ percentile median separation {\reffninety} is about $4-4\.5$ 
times  $r_\perp(z_0),$ depending on $z_0$ and 
cosmological model to no more than about
$10\%.$ Again, the relative robustness of {\reff} to $\epsilon$ 
implies that a constant value of {\reffninety} in this range can 
be used for practical purposes without integration of Limber's Equation
being required.

The dependence of {\reff} on the shape 
of $N_z$ is not totally negligible. This can be seen to some extent 
in Fig.~\ref{f-epsz0}(a), which shows that for the same
{\zmed}, the values of {\reff} can differ by up to about a factor of two.

To confirm this more explicitly, various published photometric redshift 
distributions estimated for the HDF have been adopted in place of that 
of \cite{Sawi97}~(1997) and the resulting dependence of 
{\reff} on $\epsilon$ for these $N_z$ estimates 
are presented in Fig.~\ref{f-shape}. The $N_z$ estimates adopted are 
shown in Fig.~\ref{f-dNdzHDF}. 

The results are similar to those for \cite{Sawi97}'s (1997) $N_z,$ though 
the maximum difference between photometric and analytical distributions 
increases from about a factor of two to a factor of three to four 
for {\reffninety} in the case of high $\epsilon$ values. The bimodality
of \cite{GwHart96}'s~(1996) $N_z$ and the high resolution of 
\cite{Mob96}'s~(1996) $N_z$ make this hardly very surprising. 

The large differences between the photometric and analytical distributions 
for \cite{GwHart96}'s (1996) $N_z$ are mainly due to the low $z$ peak. Obviously 
there is uncertainty in the correctness of photometric redshifts, but 
if $N_z$ (HDF) really was as bimodal as \cite{GwHart96} (1996) estimated with 
a strong low $z$ peak, then not only would many galaxy pairs which 
contribute to $w_0$ be closer to one another
than would be expected from an analytical unimodal distribution, 
but a large part of the $w_0$ signal would be from non-cosmological 
distances, at which $r_\perp \sim 10h^{-1}$~kpc. In this case, the 
meaning of a $w_0$ estimate would have to be analysed in considerable 
detail, or else redshift separation would have to be used to 
estimate $w(r_p)$ (\cite{DavP83}~1983) rather than $w_0$ in order
to extract quantities with simple physical meaning.

In summary, the earlier photometric redshift distributions to that deduced 
by \cite{Sawi97}~(1997) imply similar values of {\reff} to those of 
\cite{Sawi97} to within much less than an order of magnitude.

It should also be noted that changing the limiting magnitude of the survey
or changing the value of $\beta$ is equivalent to a change in {\zmed}
and/or in the shape of $N_z.$

\subsection{Dependence on Other Parameters} \label{s-otherdep}

Since $r(z,u)$ is dominated by the
perpendicular separation $r(z,u=0)=\theta\, \dprop (z)$ 
(for $\theta\ll 1$~rad), 
{\reff} is nearly 
proportional to $\theta.$ Indeed, although the line-of-sight
galaxy separation for a given $z$ is not directly proportional to 
$\theta$ [i.e., $r(z,u,\theta_2)/r(z,u,\theta_1)\neq \theta_2/\theta_1$ 
if $u\neq 0$], the required domain in $u$ to obtain 
a fraction $p$ of the integral
of $w(\theta,m)$ expands sufficiently (but remains small enough) 
that the value of {\reff} remains proportional to
$\theta$ to the precision of the calculation 
(for $p=50\%$ and $p=90\%$). 

Values of {\refffifty} and {\reffninety} for the nominal angle 
of \cite{Vill96}'s (1996) $w$ determination 
[$w(\theta,m)$ is normally interpolated or extrapolated to the same angle
for a large number of different surveys in order to compare the
relative amplitudes of the surveys], i.e., $\theta=$~{1\arcsec}, 
imply {\reff} values four times lower than those displayed in 
Fig.~\ref{f-epsz0}, i.e., {\refffifty}~$\sim 10h^{-1}$~kpc and
{\reffninety}~$\sim 40h^{-1}$~kpc. While such a nominal angle is
primarily meant for comparison of amplitudes of $w,$ the physical
meaningfulness or otherwise should obviously be kept in mind 
when examining such 
figures!

Brighter faint galaxy surveys 
having most of their signal at, say, {40\arcsec},
would have their {\reff} values multiplied by ten, i.e., 
{\refffifty}~$\sim 500h^{-1}$~kpc and
{\reffninety}~$\sim 1500h^{-1}$~kpc. 

The effects of cosmological parameters on {\reff} are shown in 
panels (a) and (c) of Fig.~\ref{f-epsz0}: 
since $d \dprop /dz$ 
is successively larger for hyperbolic and cosmological 
constant dominated, flat cosmologies, the values of $r(z_0,0)$ and
consequently {\reff} increase respectively. The increase in $r(z_0,0)$ 
is in fact faster than that of {\reff}, so that Fig.~\ref{f-epsz0}(b)
shows a slight reversal in this dependence for the ratio
{\reff}$/r(z_0,0),$ though the main effect is in $r(z_0,0).$

\section{DISCUSSION} \label{s-discuss}

The results above can be summarised as order of magnitude estimates: 
\begin{eqnarray}
\mbox{\reff} &=& (1+\Delta_p)\; r_\perp[\mbox{\zmed}(N_z)]
\label{e-reff}
\end{eqnarray}
where
\begin{eqnarray}
r_\perp[\mbox{\zmed}(N_z)] &\equiv& r[z=\mbox{\zmed}(N_z),u=0,
	\Omega_0=1, \nonumber \\
	&& \lambda_0=0]  \nonumber \\
	&=& \theta\,\dprop[\mbox{\zmed}(N_z),\Omega_0=1,\lambda_0=0] 
		\nonumber \\
	&=& {2c \over H_0} \,\theta
	\left( 1 - {1 \over \sqrt{1+\mbox{\zmed}(N_z)}} \right)
	\nonumber \\
&&
\label{e-reffrp}
\end{eqnarray}
and 
\begin{eqnarray}
1 \ltapprox \Delta_{50} \ltapprox 2 \nonumber \\
2 \ltapprox \Delta_{90} \ltapprox 7
\label{e-reffdel}
\end{eqnarray}
for $\gamma=-1\.8,$ $-1\.2 < \epsilon < 3,$ $0\.1< \Omega_0 <1\.0,$ 
$0\.0 < \lambda_0 < 1-\Omega_0,$ $1 < \mbox{\zmed}(N_z) < 2\.5$ 
and for $N_z$ shapes as ``smooth'' as in Eq.~\ref{e-smoodndz} or
no ``rougher'' than that of \cite{Sawi97}'s~(1997) redshift distribution.

These relations enable {\reff} to be estimated for observational
angular correlation function measurements.
Many of the brighter ($B\ltapprox 24,$ $V\ltapprox 25$ or $R\ltapprox 26$) 
faint galaxy angular correlation measurements 
(\cite{Neu}~1991; \cite{PI92}~1992; \cite{Couch93}~1993;
\cite{IP95}~1995) show significant power law correlations on scales ranging 
from $\sim 10\arcsec$ to $\sim 10\arcmin.$ 
Adopting {\zmed}~$\approx 0\.5$ as a typical value for the above surveys, 
which should give $r_\perp$ to be correct within better than an order of
magnitude, Eqs~\ref{e-reff}-\ref{e-reffdel} (which remain valid under 
extrapolation to {\zmed}~$=0\.5$)
imply that 
$50h^{-1}$~kpc $\ltapprox$ {\refffifty} $\ltapprox 3h^{-1}$~Mpc and 
$200h^{-1}$~kpc $\ltapprox$ {\reffninety} $\ltapprox 12h^{-1}$~Mpc
for these angular ranges. Spatial correlations are well established at
most of these spatial separations, so interpretation in terms
of Eq.~\ref{e-defneps} is likely to be a good first approximation.

Possibly of interest in these cases is the higher end of the length scales.
\cite{Love92}~(1992) only consider their power law fit to $\xi(r)$ to
be good to about $20h^{-1}$~kpc. The surveys covering the largest solid
angles may well be affected by---or be used to estimate---spatial 
correlations at separations difficult to measure by local surveys such
as the Stromlo-APM survey of \cite{Love92} 
Indeed, \cite{IP95}~(1995) discuss
the degree to which their data is affected by the large scale behaviour 
of $\xi(r).$ Their angular correlation functions extend to about $0\.4\deg.$
The analysis here implies that at $0\.4\deg,$ 
{\reffninety}~$\sim 30h^{-1}$~Mpc, so careful analysis of this data might
be able to produce estimates of the perturbation power spectrum on scales
comparable to that in the Stromlo-APM survey.

The brighter angular correlation measurements on smaller fields
(e.g., \cite{Ef91}~1991; \cite{RP94}~1994; \cite{BSM95}~1995)
cover smaller angular ranges, from $\sim 10\arcsec$ to $\sim 1\arcmin,$
so again adopting {\zmed}~$\approx 0\.5$ implies that 
$50h^{-1}$~kpc $\ltapprox$~{\refffifty}
$<$~{\reffninety} $\ltapprox 1.2h^{-1}$~Mpc for these observations.
This is again within typical estimates of power law behaviour for 
$\xi(r),$ e.g., \cite{DavP83}~(1983) and \cite{Love92}~(1992), though
$50h^{-1}$~kpc is already smaller than typical estimates of the radii
of the haloes of $L^*$ (bright) galaxies. 

Raising the estimate of {\zmed} from 0.5 
to {\zmed}~$=1\.2$ (the value for \cite{Sawi97}'s HDF analysis) would
double each of these separation sizes. Alternatively, conversion to 
proper units, which are relevant if considering galaxy haloes at $z\approx1$
to be collapsed and dynamically stable objects, requires division by
$(1+z)\approx 2,$ i.e., halving the separation estimates.

This brings us back to the worry raised above---that spatial 
correlations are being measured
for galaxies which are closer together than the radii of their parent haloes!
However, since a large part of the signal in the cases 
just-mentioned comes from
larger separations, this should not affect the observational validity
of using Eq.~\ref{e-defneps} as a first order estimator.

The reliance upon small separations becomes a stronger concern for the
angular correlation function of the HDF galaxies 
(and for possible future estimates of $w(\theta,m)$ using
the HST, the Next Generation Space Telescope, or if adaptive optics
can be used for faint fields using ground-based telescopes).

The $I_{AB}<27$ limit of 
\cite{Sawi97}'s~(1996) analysis of the HDF galaxies
corresponds to $I\ltapprox 26\.5$ 
(in the standard Vega-based system), and the range of the galaxies'
colours (e.g., \cite{Metc96}~1996) is approximately 
$0 \ltapprox R-I\ltapprox 2.$ 
Hence, the redshifts appropriate for 
\cite{Vill96}'s (1996) $R$-band magnitude ranges 
$R<26\.5$ to $R<28\.5$ used in estimating $w(\theta,m)$ should be similar
to those in \cite{Sawi97}'s analysis, though the bluest galaxies not present
in \cite{Sawi97}'s (1997) analysis could be expected to be at somewhat 
lower redshifts. If we consider the range of significant signal
in \cite{Vill96}'s (1996) estimate to be from $2\.5\arcsec$ to $10\arcsec,$
this then implies for {\zmed}~$\approx1\.2$ that
$25h^{-1}$~kpc $\ltapprox$~{\refffifty}
$<$~{\reffninety} $\ltapprox 400h^{-1}$~kpc for these observations.

In this case, almost none of the signal comes from separations for 
which \cite{Love92}~(1992) found significant correlations fit 
by a power law, and the similarity to halo sizes is significant.
Possibly one of the neatest measurements 
of halo extent is by Lyman-$\alpha$ and metal line 
absorption systems in front of quasars. For instance, 
\cite{BeBo91}~(1991), 
\cite{Bech94}~(1994), 
\cite{Lanz95}~(1995), 
\cite{FDCB96}~(1996) and \cite{LeBrun96}~(1996)
estimate gaseous halo radii for both sorts of systems
of around $50-200h^{-1}$~kpc. This range
matches very closely the range in {\reff} just mentioned. 

[However, it should be noted that (1) the Lyman-$\alpha$ systems 
may not be associated with galaxies (e.g., \cite{LeBrun96}~1996) 
and (2) there is marginal rotation curve 
evidence that the halo of the Galaxy is no
more than 15~kpc in radius (\cite{HonSof96}~1996, 
but see \cite{BD97}~1997), implying a mass for 
the Galaxy about an order of magnitude smaller than most estimates.]

Is the value of {\zmed} adopted here too low? While \cite{Sawi97}'s (1997)  
photometric redshift analysis uses template spectra which account
for both internal reddening and high-$z$ Lyman absorption, earlier 
analyses, e.g., \cite{Mob96}~(1996), suggested that {\zmed} should
be as high as {\zmed}~$=2\.1.$
Even if the redshifts of the $R$-selected sample contributing to
\cite{Vill96}'s $w(\theta,m)$ estimates (or to independent estimates
of, say, an $I$-selected sample) 
were as high as $z_0\approx 2\.1,$ these effective separations would
only increase by about $30\%-50\%$ (depending on cosmology), according
to Eqs~\ref{e-reff}-\ref{e-reffdel}. The spatial separations would
still correspond to typical estimates of galaxy halo radii, particularly
when applying the conversion from comoving to proper coordinates.
Note that the values adopted for use 
in Eq.~\ref{e-smoodndz} in \cite{Vill96}'s (1996) models shown in their 
Fig.~2 for $R<26\.5$ to $R<28\.5$ are $1\.4\ltapprox z_0\ltapprox 1\.8.$ 

\nice \tabColl ::::

This result brings to mind the suggestions, based on 
morphological analysis of the HST Medium Deep Survey (MDS) and 
the HDF 
(\cite{Grif94}~1994; 
\cite{Caser95}~1995; \cite{Driv95a}~1995a,b; 
\cite{Glaz95}~1995; \cite{Abrah96}~1996;
\cite{vdB96}~1996),
 that many of the HDF galaxies are the ``building blocks''
of future galaxies which later merge together. 
(\cite{Pasc96} (1996) make a similar suggestion based on 
HST imaging and Multiple Mirror Telescope spectroscopic redshifts.)
In this case, the HDF galaxies' haloes 
should be smaller than present day galaxy haloes, and so the
visible galaxies could be correlated without their haloes necessarily
overlapping. 

However, \cite{DavP83}'s (1983) observation of correlated galaxies down
to $10h^{-1}$~kpc is a local observation, it is not an observation of 
primordial galaxies. It is quite noisy, so perhaps is already affected
by the detailed nature of the galaxy-halo relationship. In either case,
the angular correlation function of the HDF must  be affected by,
and hence offer some clues to, the relationship between galaxies and
haloes.

A complementary analysis to that presented here is that of 
\cite{Colley96}~(1996; 1997). \cite{Colley96}~(1996)  
 measured a very strong
angular correlation for ``galaxies'' selected by colour to be at high
redshifts ($z\gtapprox 2$) and point out that the detected objects
may in fact be star-forming regions within ``normal'' galaxies,
dimmed only by $(1+z)^2$ in surface brightness rather than $(1+z)^4$ 
since they are (nearly) unresolved by the HST. In \cite{Colley97}~(1997),
apart from a dynamical discussion of different scenarios, the authors add a 
nearest-neighbour analysis which strengthens their argument that 
there is an excess of close objects separated by less than about 
$1\arcsec.$

The result of the effective separation calculations presented here
only strengthens \cite{Colley96}'s model. As they point out in their
conclusions (\cite{Colley97}~1997), two galaxies at identical redshifts 
separated by $1\arcsec$ at the redshift range of their sample are
separated by roughly 6~kpc in proper units. 
Since {\reff}~$\sim r_\perp,$ if the objects 
were indeed individual galaxies, and if a much larger field were measured
in order to reliably measure $w(\theta,m)$ at these small angles and
faint magnitudes, then about 50\% of the spatial correlation signal 
would be from galaxies separated by less than 
about $6h^{-1}$~kpc, and 90\% of the signal for galaxies closer to one
another than about $24h^{-1}$~kpc. These values, in both comoving and
proper coordinates in order to facilitate deductions from both cosmological 
and galaxy formation points of view, are presented in 
Table~\ref{t-Colley}. 

Returning to the full sample of the HDF galaxies, 
for which $25h^{-1}$~kpc $\ltapprox$~{\refffifty}
$<$~{\reffninety} $\ltapprox 400h^{-1}$~kpc,  
are there any justified alternatives to the extension of 
Eq.~\ref{e-defneps} to these separations, either
observationally or theoretically? 

A comparison of the theoretical evolution
of the spatial correlation function to \cite{Vill96}'s (1996) HDF estimates
is given by \cite{Matar96}~(1996), for 
an $\Omega_0=1, \lambda_0=0$ CDM universe where the full linear and non-linear
evolution is given by a formula fit to gravity-only N-body simulations 
(\cite{Ham91}~1991; \cite{JMW95}~1995). \cite{Matar96}'s (1996) Fig.~5 
(for no bias factor) is approximately 
consistent with the HDF correlations, whilst
\cite{Vill96} (1996) found that a spatial correlation growth rate
of $\epsilon=0\.8$ in Eq.~\ref{e-defneps} 
was not sufficient to provide low enough correlations
unless the value of $r_0$ is decreased, i.e., that a transition in populations
is hypothesised. Differences at the sub-Mpc level could explain this
better fit. However, the CDM simulations used for \cite{JMW95}'s 
formula have only 1 particle/(350$h^{-1}$~kpc)$^3$ and a force resolution of
32~$h^{-1}$~kpc, so may not be valid at these scales. Additionally,
of course, they cannot address 
the question of how to relate stellar galaxies to dark matter 
dominated haloes---unless this is governed by gravity alone, which seems
unlikely.

An empirical, but theoretically motivated, alternative
would be \cite{Peac96}'s (1996) analysis based on a two-power law fit
to galaxy correlations of the APM (\cite{Madd90w}~1990a; 1990b) and IRAS 
(\cite{IRAS92}~1992) surveys. However, \cite{Peac96} 
finds that this fit implies that 
for an $\Omega_0=1, \lambda_0=0$ universe, an extra
correlation at scales of $r \ltapprox 2\pi h^{-1}$~Mpc is needed to
fit the CFRS (\cite{CFRS-VIII}~1996) data, unless a scale-dependent
bias factor is used in combination with a constant bias factor. 
\cite{Matar96}'s (1996) use of \cite{JMW95}'s (1995) 
 N-body motivated $\xi(r,z)$ description better fits the CFRS data
(for no bias), so would maybe more appropriate for an 
extension of the present work.

An alternative empirical improvement to Eq.~\ref{e-defneps} could be to
use the $w$ estimates of \cite{Inf96}~(1996) on very small scales 
(down to $\approx 1\arcsec$) in 
a survey covering a very large solid angle ($2\.3$~sq.deg.), hence, having 
enough precision to make such estimates. \cite{Inf96}~(1996) find that
at the smallest separations, 
$2\arcsec \ltapprox \theta 
\ltapprox 6\arcsec$, there is a correlation signal about a factor
of three higher than 
that of a $\gamma=1\.8$ power law extrapolated from larger angles. 
To the extent that the above {\reff} calculations can 
be extrapolated to this ``highly non-linear'' case, 
the value $r_\perp=24h^{-1}$~kpc (as defined in Eq.~\ref{e-reffrp})
implies that half of this signal comes from 
three-dimensional galaxy pair separations of about this size, and 40\%
more comes from separations up to about $100-200h^{-1}$~kpc. 
(Note that \cite{Inf96} adopt proper units. The conversion follows from
using {\zmed}~$=0\.35.$)
The values
of $\xi(r,z)$ for $r \ltapprox 20h^{-1}$~kpc, without contamination from
``normal'' correlations at larger angles, could therefore be even higher 
than might be expected from simple inspection of \cite{Inf96}'s (1996) figures.

\section{CONCLUSIONS}\label {s-conclu}

In order to see what length scales and redshifts correspond to 
the spatial correlations $\xi(r,z)$ represented
in recent measurements of the faint galaxy angular correlation function 
$w(\theta,m),$ 
Limber's Equation (Eq.~\ref{e-limber}) has been examined over
a range of likely possibilities for $\xi$ and $N_z(z,m).$ The separation into
$N_z(z,m)$ and $\xi[r(z,u),z]$ and the contours in the $z-u$ plane
which contribute to the integral have been shown, revealing that 
observational $w$ measurements combine spatial correlations from a
wide range in mean redshift.

A simple definition of an effective separation {\reff} to quantify the
relevant length scales for $w$ 
has been defined as the median (comoving) separation
within the contour of constant $(\partial N/\partial z)^2 \xi[r(z,u),z]$ 
contributing a fraction $p$ of the numerator of Eq.~\protect\ref{e-limber}. 
This is similar in order of magnitude to 
the perpendicular distance $r_\perp\equiv \dprop \theta$ at
the characteristic redshift, $z_0$, 
of simple analytical redshift distributions
(Eq.~\ref{e-smoodndz}), and hence to 
the median redshifts of such distributions.

More precisely,
for the parameter space covered, {\refffifty} is about $5\%-15\%$
greater than $r_\perp(z_0),$ i.e., about $10\%-20\%$ 
greater than $r_\perp(${\zmed}$)$ (for $\beta=2\.5$ in 
Eq.~\ref{e-smoodndz}); while {\reffninety} is about $4-4\.5$ 
times  $r_\perp(z_0).$ 

For a more realistic redshift distribution, e.g., the photometric
redshift distribution of \cite{Sawi97} (1997),  
{\reff} may be different by up to about a factor of two to the {\reff} values 
for the analytical redshift distributions if $-1\.2 < \epsilon < 3.$

In summary, values of {\reff} for typical median redshifts of most faint
galaxy angular correlation measurements indicate that the use 
of Eq.~\ref{e-defneps} does not imply an extrapolation (in galaxy 
pair separation) 
much below or above 
observational spatial correlation measurements. (The validity of redshift 
dependence scaling by $(1+z)^{-(3+\epsilon-\gamma)}$ is not tested, however.)

The scales on which HDF galaxies are correlated (as measured by
\cite{Vill96}~1996) imply comoving separations of
$25h^{-1}$~kpc $\ltapprox$~{\refffifty}
$<$~{\reffninety} $\ltapprox 400h^{-1}$~kpc. That is, for typical
redshifts over $z\gtapprox1,$ a substantial fraction of the angular 
correlation signal is generated by galaxies spatially separated, in proper 
units, by around $10-100h^{-1}$~kpc. While spatial correlations of galaxies
at these separations have been observed for local galaxies 
(\cite{DavP83}~1983), this is a scale at which 
halo and/or galaxy existence, let alone interactions, might 
strongly modify the spatial correlation function.

\medskip
We wish to dedicate this paper to the memory of the
late Prof. Roger Tayler, whose support for this collaboration was
well appreciated. We thank Alain Blanchard for useful discussion on 
this topic and B. Mobasher for providing tables in electronic format.
BFR acknowledges a Centre of Excellence 
Visiting Fellowship (National Astronomical Observatory of Japan, Mitaka), 
support from PPARC and from Starlink computing resources. 

\subm \clearpage ::::
\subm \tabColl ::::

\subm \clearpage ::::

\subm \clearpage ::::

\subm \fperc ::::
\subm \fepsz0 ::::
\subm \fshape ::::
\subm \fdndz ::::

\end{document}